\documentclass[conference]{IEEEtran}
\IEEEoverridecommandlockouts
\usepackage{cite}
\usepackage{amsmath,amssymb,amsfonts}
\usepackage{amsthm}
\usepackage{float}
\usepackage{comment}
\usepackage{algorithmic}
\usepackage{graphicx}
\usepackage{textcomp}
\usepackage{xcolor}
\usepackage{balance}
\usepackage{hyperref}
\usepackage{subcaption}
\def\BibTeX{{\rm B\kern-.05em{\sc i\kern-.025em b}\kern-.08em
    T\kern-.1667em\lower.7ex\hbox{E}\kern-.125emX}}

\theoremstyle{definition}

\theoremstyle{definition}

\begin{document}

\title{LEAST: a Low-Energy Adaptive Scalable Tree-based routing protocol for Wireless Sensor Networks}

\author{\IEEEauthorblockN{Amirmohammad Farzaneh\IEEEauthorrefmark{1},
Mihai-Alin Badiu\IEEEauthorrefmark{2}, and Justin P. Coon\IEEEauthorrefmark{3}}
\IEEEauthorblockA{Department of Engineering Science,
University of Oxford\\
Oxford, UK\\
Email: \IEEEauthorrefmark{1}amirmohammad.farzaneh@eng.ox.ac.uk,
\IEEEauthorrefmark{2}mihai.badiu@eng.ox.ac.uk,
\IEEEauthorrefmark{3}justin.coon@eng.ox.ac.uk}}

\maketitle

\begin{abstract}
Routing is one of the critical and ongoing challenges in Wireless Sensor Networks. The main challenge has always been to have a routing protocol that reduces the communication overhead, hence saving the energy of the sensors in the network. Hierarchical routing protocols are known to be the most energy-efficient routing protocols for Wireless Sensor Networks. In this paper, a more generalized hierarchical routing protocol is introduced for Wireless Sensor Network, which is based on tree data structures. The clustering in the proposed protocol has the format of a general tree, which is constructed in an adaptive manner based on the distance of the sensors. Results show that the proposed tree-based protocol introduces a significant benefit in energy consumption and lifetime of the network over the existing hierarchical approaches.
\end{abstract}

\begin{IEEEkeywords}
Wireless Sensor Network, routing, trees
\end{IEEEkeywords}

\section{Introduction}

Wireless Sensor Networks (WSNs) are formed of a number of sensor nodes in an area, which gather and collect information about the environment in which they are located. This information is then collected by a special node in the network, often known as the Base Station (BS) or Sink Node (SN). The BS is the gateway of communicating the information collected by the sensors to the outside world, via the internet. WSNs can vary in size anywhere from very small local networks, to networks with tens of thousands of sensor nodes in some practical scenarios. The sensors can be mobile or stationary. However, they are usually stationary or have very limited movement in practice. It is usually assumed that all the nodes in a WSN, other than the BS, have limited power supply. They are powered with a battery, installed at a location, and barely ever looked after. This is because constant maintenance of these sensors would be costly. Therefore, power consumption of the sensor nodes is one of the main challenges in WSNs.

Routing protocols form an inseparable part of any network in general. They are also of extreme importance when it comes to WSNs. The reason for this is that these protocols need to be designed in a way that use the least computation and communication power as possible. Additionally, the rapid growth and fast deployment of WSNs demands a more scalable approach to the problem of routing in such networks. Another important point of routing protocols for WSNs is their data-centric nature, rather than address-centric \cite{obaidat2014principles}. This means that data can be collected and processed locally, and then be sent to the BS.  As a result of all these factors, designing routing protocols for WSNs is still an ongoing area of research.

The research conducted so far on developing routing protocols for WSNs is very vast. Some of the most notable ones are SPIN \cite{heinzelman1999adaptive, kulik2002negotiation}, Directed Diffusion \cite{intanagonwiwat2000directed}, and ACQUIRE \cite{sadagopan2003acquire}. However, the class of hierarchial routing protocols for WSNs seems to be attracting more attention in the recent years. This category of routing protocols is known to be very energy efficient and scalable \cite{al2004routing}. In fact, some sources consider hierarchical routing protocols to be the most energy efficient routing protocols for WSNs \cite{bhushan2019routing}.

LEACH (Low-Energy Adaptive Clustering Hierarchy) \cite{heinzelman2002application} is perhaps the most well-known and widely-used hierarchical routing protocol for WSNs. After LEACH, numerous extensions to it were introduced that tried to overcome the limitations of LEACH. PEGASIS \cite{lindsey2002pegasis}, TEEN \cite{manjeshwar2001teen}, and HPAR \cite{li2001hierarchical} are among other hierarchical routing protocols. Most of the existing hierarchical routing protocols provide a three-level hierarchy. A number of nodes at each round are chosen as Cluster Heads (CH), and other nodes get assigned to one of these cluster heads. This will essentially create a rooted tree structure, with the BS acting as the root, CHs acting as the children of the root, and other nodes are children of a CH. However, we believe that extending this tree to a multi-level tree can significantly improve energy consumption. This is because of the fact that the transmission energy required to send a packet over a wireless channel between two nodes with a distance of $d$ is proportional to $d^2$, which implies that breaking down the distance into multiple smaller steps can reduce energy consumption. There exist multi-level, or tree-based, hierarchical routing protocols for WSNs, but each of them have their own limitations. For instance, in a multi-level LEACH protocol \cite{kodali2014multi}, the number of levels are fixed, and the flexibility of the routing topology is therefore reduced. Another example is EADAT (Energy-Aware Data Aggregation Tree) \cite{ding2003aggregation}, which takes into account the residual energy of nodes to construct routing paths. However, it has been seen that this method is not necessarily energy-efficient in the end, as it does not take into account the distance between nodes. 

In this paper, we introduce LEAST, a Low-Energy, Adaptive, Scalable, and Tree-based routing protocol for WSNs. The routing paths in LEAST essentially form a rooted tree, with the BS acting as the root node. The structure of the tree is formed based on fairness in dividing the consumed communication power between nodes, while minimizing the distance between hops. The assumptions about the sensor network and some of the notations used throughout the paper are described first. Then, LEAST is introduced and the different steps of the protocol are described in detail. It will be shown how adjusting different parameters of LEAST can be used to tailor the protocol for different networks. The performance of LEAST is then analysed and compared with LEACH. Afterwards, simulation results of LEAST are exhibited, which show a significant improvement over existing hierarchical routing protocols. Ultimately, the paper is concluded and potential ways that LEAST can be improved are mentioned and discussed.

\section{Preliminaries and network model}

We assume the network to have $n$ sensor nodes, and one Base Station. We also assume all the sensors to have enough transmission power to be able to communicate with any other node in the network if needed. We consider a TDMA/CDMA MAC protocol, such as the one described in \cite{ma2014hybrid}, to avoid collisions in the communication between nodes. Sensor nodes in the network can start with arbitrary energy levels, as energy levels are not considered in this version of LEAST. It is also assumed that most of the communication in the network will take place to and from the BS. If by any chance communication needs to take place between two sensor nodes in practical applications, they can do so by using the BS as an intermediary node. 

\begin{figure}
    \centering
    \includegraphics[width = 0.8\columnwidth]{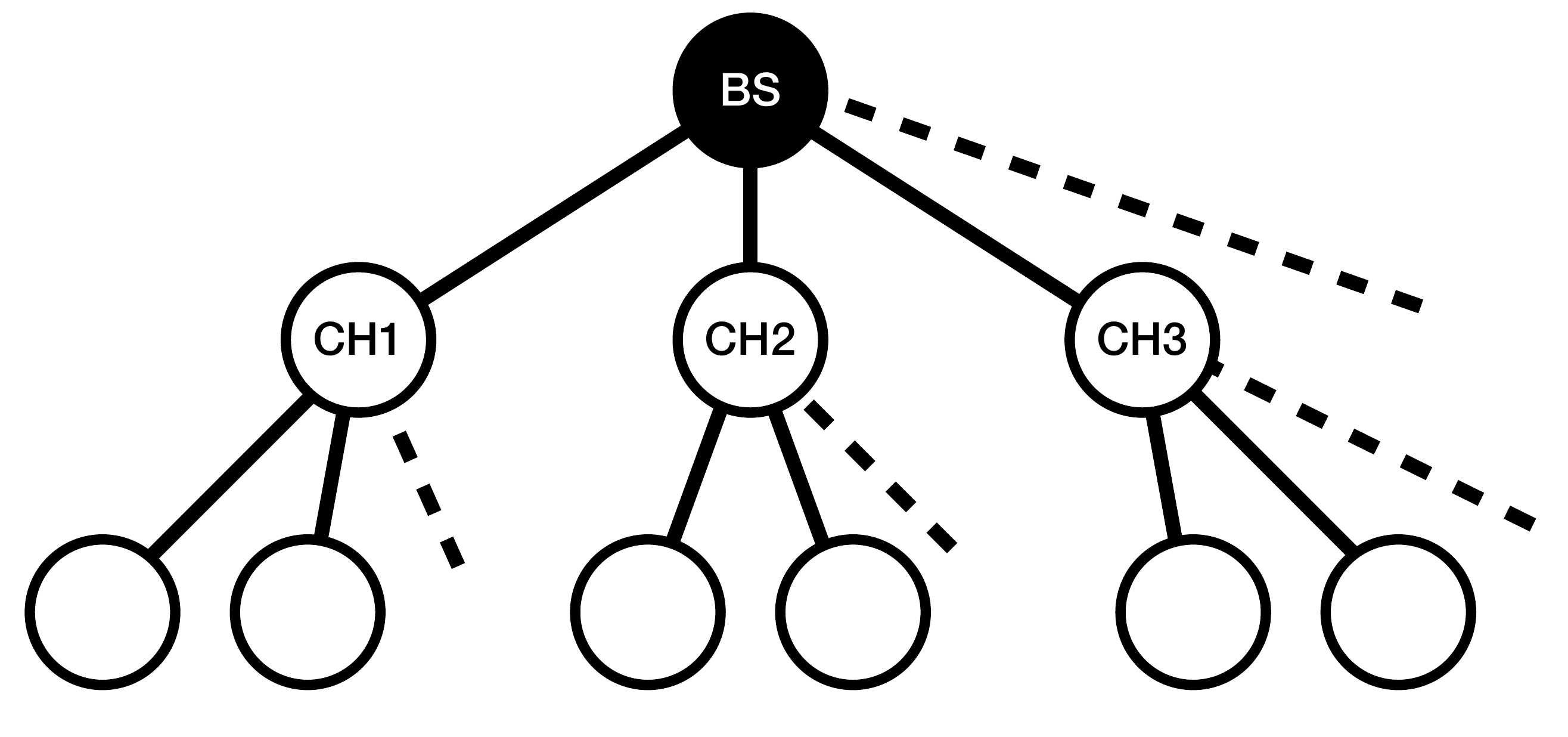}
    \caption{LEACH routing map}
    \label{level_1}
\end{figure}

For ease of deployment, LEAST is designed to be an extension of LEACH. This way, it will be easy to modify the systems that are currently based on LEACH to operate on LEAST. LEACH is divided into rounds, with each round having two phases: the setup phase and the steady-state phase. In the setup phase, the routing map is formed, and the messages are transmitted using these routes in the steady-state phase. In the setup phase, every node will elect itself as a Cluster Head (CH) with a probability of $p_{CH}$. The CHs will then be in charge of passing on the messages between the sensor nodes and the BS. Additionally, all the other nodes will need to choose a CH to be connected to. Non-CH nodes will choose the CH with the minimum distance to them as their parent. Fig. \ref{level_1} illustrates how the routing tree looks like in LEACH.

\section{Protocol description}

In this section, we describe how LEAST works, and its differences with LEACH. Just like LEACH, LEAST is divided into rounds, and each round consists of a setup phase and a steady-state phase. The first round of LEAST will be exactly like that of LEACH, during which nodes will create a three-level routing map for the network. The network map will essentially represent the path that has to be taken from nodes in the network to reach the BS. Every time it is mentioned that a message is sent from a sensor node to the BS or vice versa, it is assumed that the path that the sent packets take is the one indicated on the map of the network during that round. After the initial map is set up by the LEACH algorithm, the other rounds will have a different approach. Unlike LEACH, the routing map in LEAST can have more than three levels. We number the levels based on their distance from the BS: the BS will be level zero, the CHs will be on level one and so on. The steps below describe the operations that need to be done at the setup phase of each round after round one.

\subsection{Selection of Host Nodes}

At each round, the nodes on the first level of the network map (children of the BS) will need to be relocated. These nodes spend a lot of energy managing the packets received from their children, and performing tasks such as data fusion. Therefore, we relocate them to a lower position to keep the balance in energy consumption. For this purpose, a number of nodes are selected as Host Nodes (HN) in each round. The HNs will act as potential future parents for the first-level nodes. However, the actual relocation does not happen until later in the setup phase of the protocol.

Every node elects itself to become a Host Node (HN) for that round with a probability of $p_{HN}$. The election process will be exactly the same as electing the CHs in the LEACH protocol. Every node that has elected itself as a Host Node notifies the BS about it, which will in turn notify the first-level nodes about the list of HNs. In order to ensure that different nodes act as an HN for the same number of rounds, we use the same formula as LEACH \cite[Eq.~3]{heinzelman2002application}. Therefore, in round $r$, a node can only decide to be an HN if it has not acted as an HN in the last $np_{HN}$ rounds.

\subsection{Selection of heir nodes}

Before first-level nodes get relocated, we need to choose which nodes are going to replace them. For each first-level node, one or more of its children are going to be chosen as its heir(s). Each of these children are going to be chosen as an heir with a probability of $p_H$, independent from each other. At the end, if any of the first-level nodes did not have an heir, one of its children will be chosen as its heir in a uniform manner. Heir nodes notify their parents about the fact that they have been selected as an heir. This is done so that the parent can uniformly select one of its children as heir in case none of them were chosen as an heir.

\subsection{Relocation of first-level nodes}

The final step of the setup phase is to relocate the first-level nodes. We remove first-level nodes from their current location, and connect each one to the HN with the minimum distance from it. Then, the heir nodes are moved up and attached to the BS. The other children of the former first-level nodes then get attached to the closest heir node of their former parent. After this, the relocation process is complete.

\begin{figure}
    \centering
    \includegraphics[width = \columnwidth]{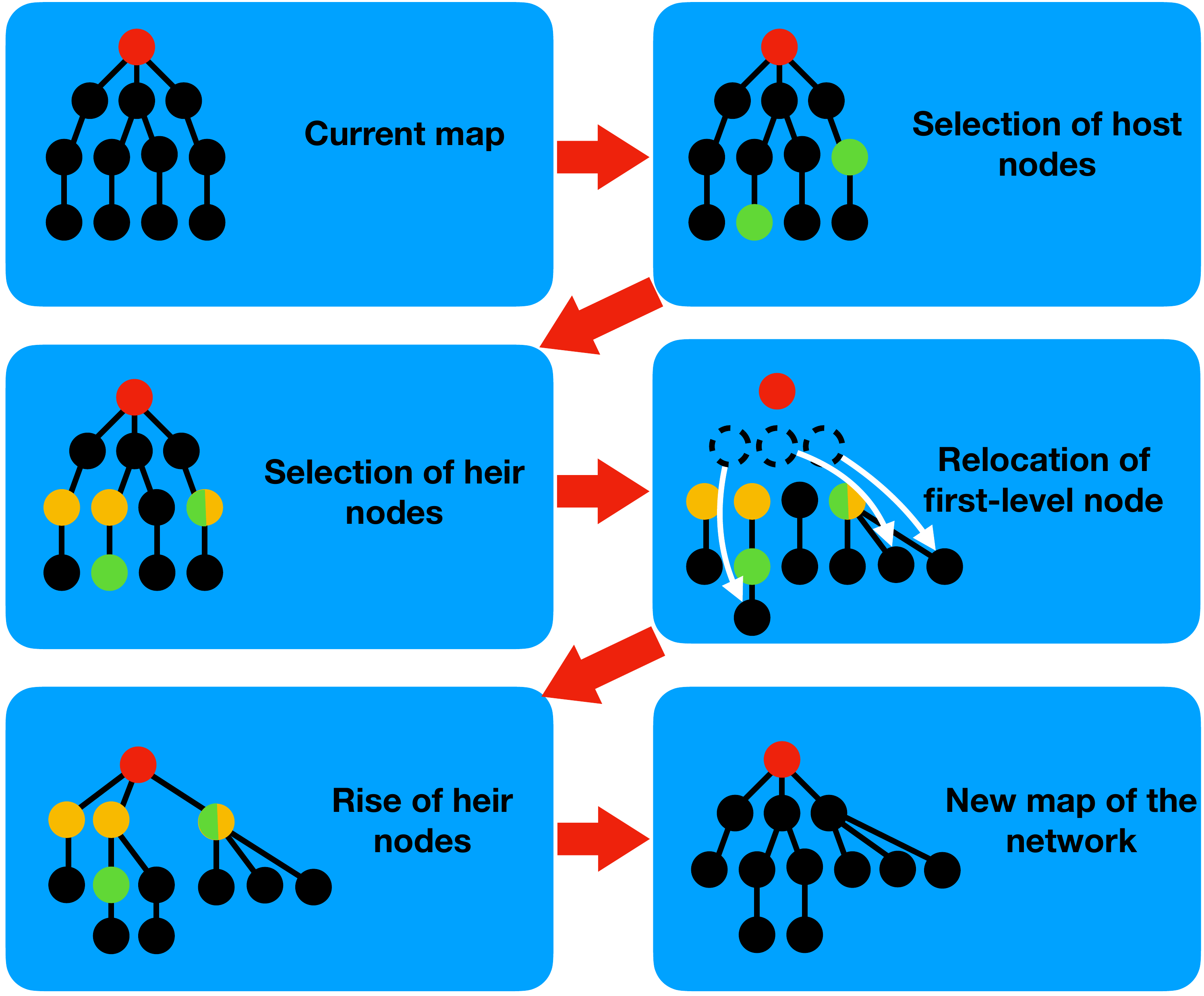}
    \caption{Setup phase of LEAST}
    \label{protocol}
\end{figure}

\begin{figure}
    \centering
    \includegraphics[width = 0.8\columnwidth]{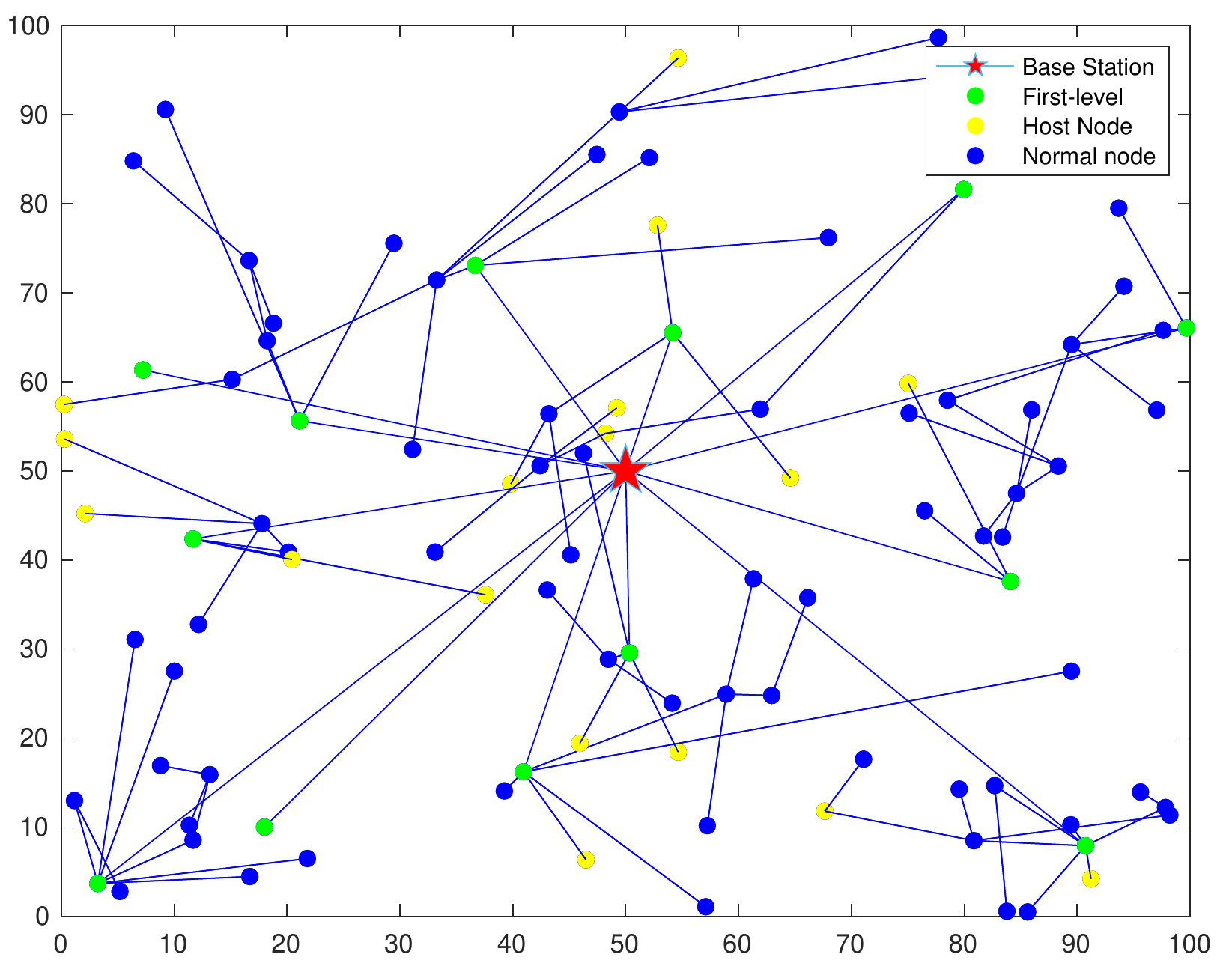}
    \caption{A snapshot of the network map for a sample network at a random around in the LEAST protocol}
    \label{snapshot}
\end{figure}

Fig. \ref{protocol} summarizes the setup phase of LEAST, at the end of which the map of the network is formed. Fig. \ref{snapshot} shows a snapshot of the network map at a random round for a sample network. It can be seen how the paths are formed in a way that is acceptable in terms of node distance, while preserving the random nature of the algorithm to distribute energy consumption fairly among nodes. After the setup phase, the steady-state phase of the round begins. It is during the steady state phase that normal communications happen between the sensors and the BS. Whenever a node wants to send a packet to the BS, it just forwards the packet to its parent, which will then forward the packet to its own parent. This process continues until the packet reaches the BS.

\subsection{Protocol parameters}

\subsubsection{$p_{CH}$}

$p_{CH}$ is a parameter that is only used once at the beginning of the protocol. This parameter indicates the proportion of nodes to be chosen as CHs to form the initial network map. During the process of node relocation, there is always at least one node replacing each first-level node. Based on the fact that initially there will be an average of $np_{CH}$ cluster heads, $p_{CH}$ actually determines the minimum number of first-level nodes in LEAST.

\subsubsection{$p_{HN}$}

$p_{HN}$ determines the proportion of the total number of nodes to be considered as the new parents of the current first-level nodes. Increasing this value will increase the number of candidates, which means that the chances of having at least one parent node with a small distance for each first-level node is increased. This would mean that the cost of communication between the current first-level nodes and their new parents can be decreased, which can contribute towards energy efficiency. However, notice that increasing $p_{HN}$ would also mean that more communication needs to happen between first-level nodes and HNs to find the closest HN to each of the first-level nodes, which will increase energy consumption. Therefore, there is a trade-off in increasing $p_{HN}$. $p_{HN}$ should therefore be chosen according to how the sensor nodes are spread and located across the network.

\subsubsection{$p_H$}

$p_H$ has been designed as a safety measure to prevent first-level nodes from having more children than expected. Increasing $p_H$ will result in more nodes in the first-level, and decreasing the number of children per first-level node. Therefore, there is again a trade-off in choosing the right value for $p_H$, which should be based on the arrangement of the sensor nodes.

\section{Energy consumption analysis}
In this section, we will try to perform a mathematical analysis of the energy consumption of LEAST, and compare it with that of LEACH. For this purpose, we assume the processing power consumed in the sensors to be negligible compared to that used for wireless communication. Additionally, we make use of the fact that energy required for wireless transmission between two nodes with a distance of $d$ is proportional to the number of packets communicated, and to $d^2$. Therefore, we will try to estimate the number of messages and the distance over which each message needs to travel during the setup stage of LEAST. Additionally, for the purpose of this analysis, we assume $p_H$ to be zero. We will discuss the effects of a nonzero $p_H$ at the end of this section.

As a detailed analysis of the transmission energy requires knowledge about the location of the sensors, or the distribution that their locations follows, we use average values for distances in this analysis. We use $\bar{d}$ to show the average distance between pairs of nodes in the network. Additionally, we use $\bar{d_m}$ to show the average value for the distance between nodes and the furthest node from them.

The first point of communication between nodes in LEAST happens when HNs have to notify first-level nodes about themselves. The most simple, but not the most efficient, way to do so is for each HN to broadcast its status. This way, the (average) consumed power for the announcement of HNs will be proportional to $np_{HN}\bar{d_m}^2$. The other phase that consumes energy is the election of heir nodes. Once these nodes are selected, they need to notify the BS, which can be done in two hops through their corresponding first-level nodes. They also need to inform their sibling nodes. This can be done with a broadcast message, but the distance average distance for this can be considered to be $\bar{d}^2$ rather than $\bar{d_m}^2$. This is because the connections to a parent are inherently formed based on closeness of the nodes, so it is very unlikely to have sibling nodes whose distance is among the largest possible pair distances in the network. Considering $p_H$ to be zero, every first-level node will only get one heir, which means that the number of first-level nodes will not change after the first round of the protocol. Therefore, we can consider the number of first-level nodes to be $np_{CH}$ on average. This would mean an average power consumption of less than $3np_{CH}\bar{d}^2$ for this phase of the round. Consequently, communication power used in LEAST can be estimated using (\ref{least_power}).

\begin{equation}
\label{least_power}
    \parbox{8em}{Average power per round of LEAST }\propto n(p_{HN}\bar{d_m}^2 + 3p_{CH}\bar{d}^2)
\end{equation}

For LEACH, each round consists of the broadcast announcement of CHs, followed by every node reaching out to its closest CH to attach to it. Using the same analysis as LEAST, we can use (\ref{leach_power}) to estimate the communication power usage of LEACH.

\begin{equation}
\label{leach_power}
    \parbox{8em}{Average power per round of LEACH }\propto n(p_{CH}\bar{d_m}^2 + (1-p_{CH})\bar{d}^2)
\end{equation}

We can now compare LEAST and LEACH using (\ref{least_power}) and (\ref{leach_power}). It can be seen that the difference between these two estimates is $n((p_{CH}-p_{HN})\bar{d_m}^2 + (1-4p_{CH})\bar{d}^2)$. Assuming $p_{CH}$ to be relatively small, and the values for $p_{CH}$ and $p_{HN}$ to be close (if not equal), it can be seen that LEACH consumes more energy compared to LEAST. This mainly comes from the fact that even though the network map is updated every round, not all nodes in LEAST need to be relocated. This saves a lot of energy in terms of the coordination that needs to be done between nodes to perform routing tasks. Additionally, note that we made some simplification assumptions in this analysis, which overestimates the power consumed in LEAST. Therefore, the actual difference between the power consumed by LEAST and LEACH is much more.

We now consider the effect of a nonzero value for $p_H$. There is always going to be at least one heir for every first-level node. Having a nonzero $p_H$
gives rise to the possibility of having multiple heirs for a first-level node. As explained earlier, $p_H$ is mainly designed as a safety measure for preventing a first-level node from having too many children, and its value is usually quite small. This might change our estimate for the messages send from the heir nodes to announce their status. However, notice that even if at one round, a first-level nodes gets multiple heirs, this will not happen again in the next few rounds as the children get divided, and the value for $p_H$ is small. Therefore, we believe that a nonzero value for $p_H$ will have a negligible effect on our estimate for the consumed power per round.

\section{Results}

In this section, we analyse the simulation results of LEAST using MATLAB. The code used for simulating LEAST is a modified version of a code designed to simulate LEACH, which can be found at \cite{leachcodematlab}.

For the simulations, the network is created in a 100x100 grid. The BS is located at the center of the grid, $(50,50)$. A number of 100 sensors are located randomly on the grid, their $x$ and $y$ coordinates are chosen uniformly. The LEAST protocol is then implemented as described in the paper. The parameters for this simulation are chosen as follows.

\[p_{CH} = 0.1, \quad p_{HN} = 0.2, \quad p_{CH} = 0.1\]

The initial energy of the sensors was set to $0.1$ Joules. Throughout the simulation, communication between sensors will reduce their energy, based on the transmission distance. A sensor node is dead if its energy level reaches zero.

We compare the performance of LEAST with LEACH. This is because of the fact that despite the numerous existing hierarchical routing protocols for WSNs, LEACH is still one of the most well-known and widely used ones. Additionally, most other protocols compare their result with LEACH, which means that comparing the performance of LEAST with LEACH will actually provide us with an image as to how LEAST will perform in comparison to other algorithms.

\begin{figure}
    \centering
    \begin{subfigure}[b]{0.8\columnwidth}
    \includegraphics[width = \columnwidth]{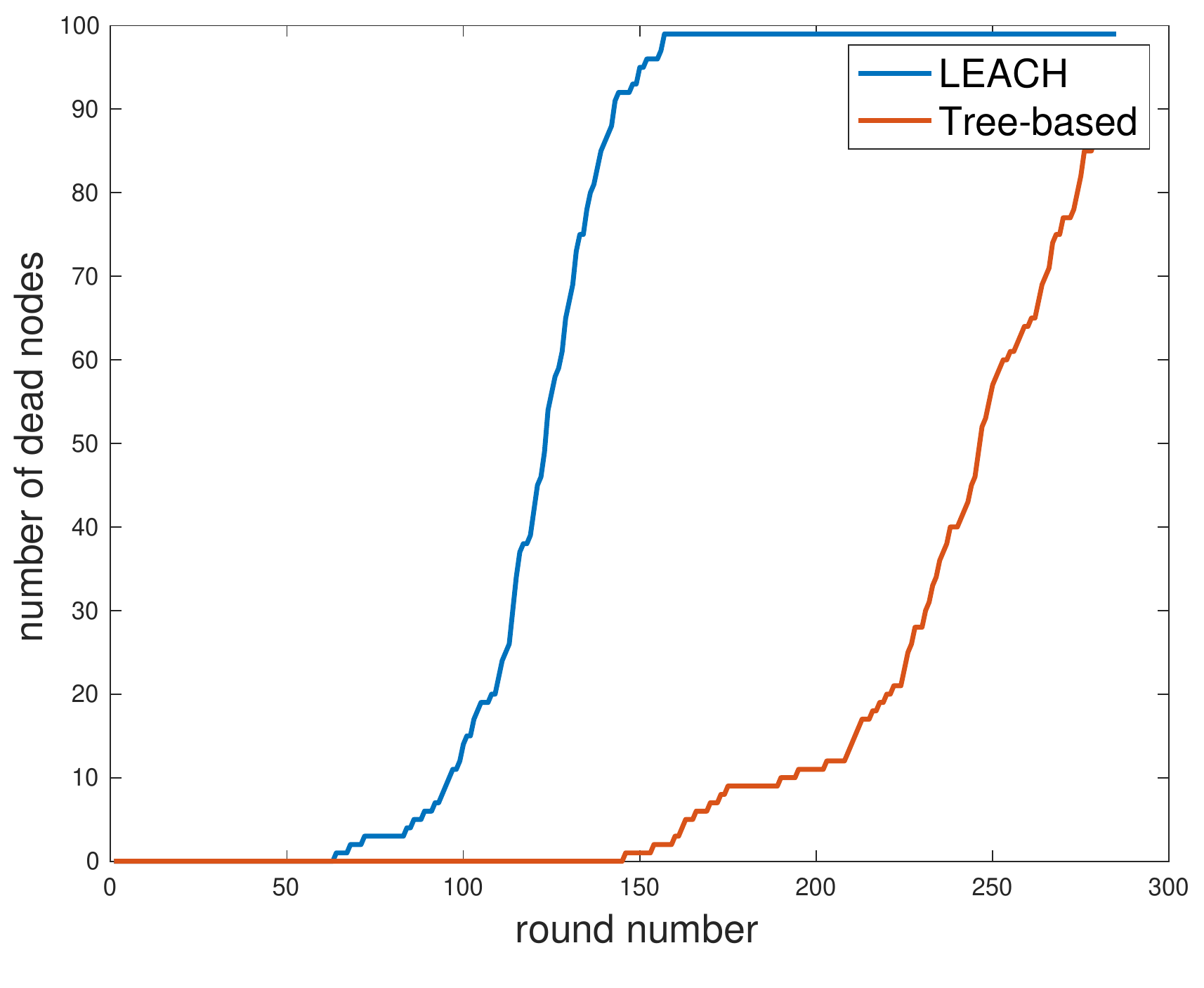}
    \caption{Number of dead nodes per round}
    \label{result1}
    \end{subfigure}
    \centering
    \begin{subfigure}[b]{0.8\columnwidth}
    \includegraphics[width = \columnwidth]{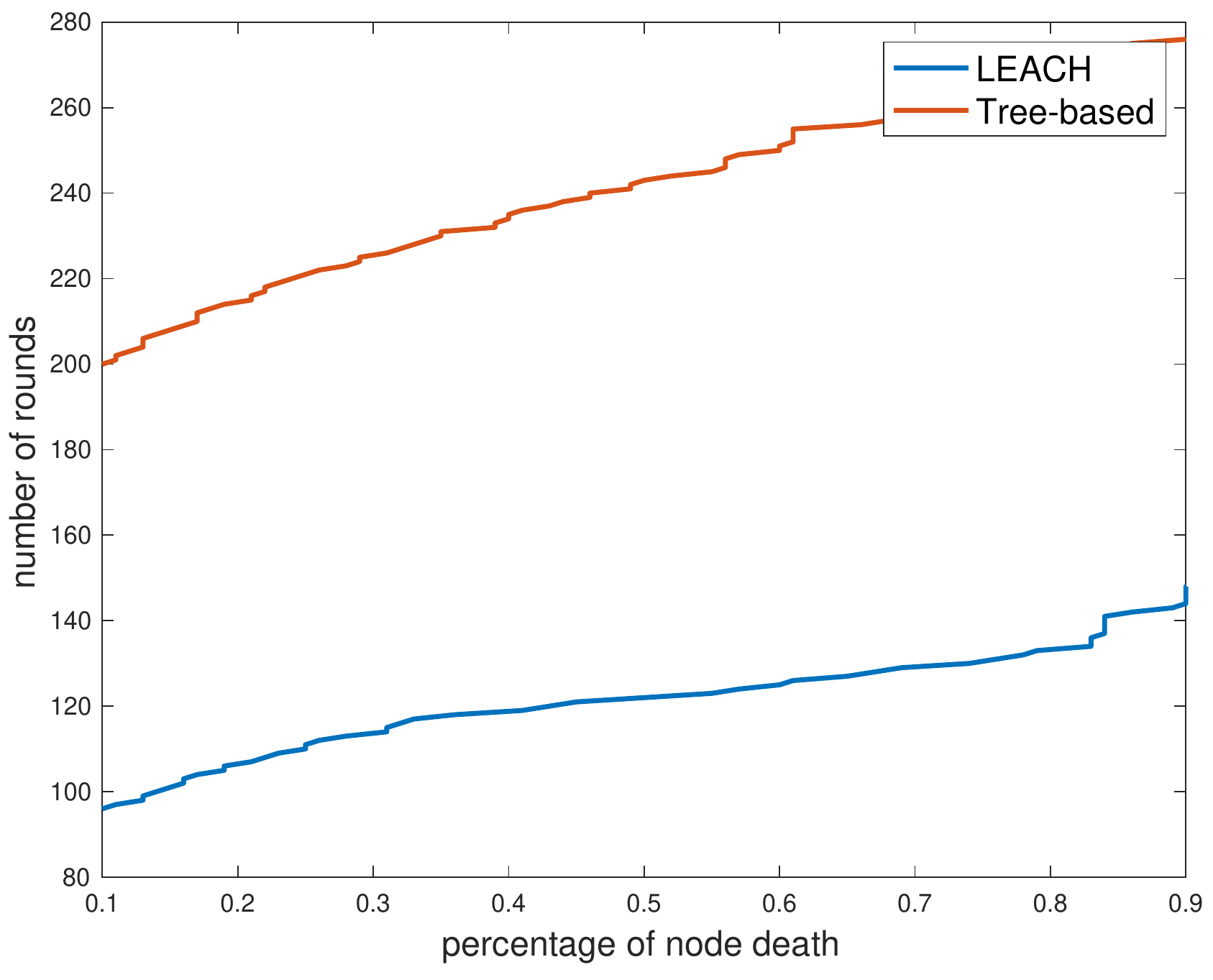}
    \caption{Number of rounds per percentage of dead nodes}
    \label{result2}
    \end{subfigure}
    \caption{Comparison of LEAST with LEACH}
    \label{dead}
\end{figure}

Fig. \ref{dead} illustrates the performance of LEAST and compares it with LEACH. The number of dead nodes in each round of the protocol for both LEAST (tree-based) and LEACH are shown in Fig. \ref{result1}. It can be seen that the first node in LEAST dies much later than LEACH. This shows a better energy performance for LEAST compared to LEACH. It can also be observed that when the nodes do start to die, LEAST exhibits a slightly lower slope, which means that nodes die in a slower pace. In order to compare LEAST with routing protocols other than LEACH, we refer to \cite{kim2005power}. \cite{kim2005power} introduces a multi-level hierarchical routing protocol for WSNs called BATR (Balanced Aggregation Tree Routing), and compares its performance with other protocols such as LEACH and direct transmission of data to the BS. \cite[Fig.~3]{kim2005power} shows this comparison, by plotting the number of rounds versus the percentage of dead nodes in each round. Fig. \ref{result2} illustrates the same plot, comparing the performance of LEAST and LEACH. Note that the scale of the vertical axis is different from \cite[Fig.~3]{kim2005power}, as the values of the starting energy of the nodes is different for these two simulations, and the exact values used in the simulations of \cite{kim2005power} were not given. However, by comparing the overall shape of Fig. \ref{result2} with \cite[Fig.~3]{kim2005power}, it can be seen that LEAST is introducing a significant improvement over all the protocols studied in \cite{kim2005power}. It can be observed that the curve for LEACH starts much higher than that of LEAST, and by comparison BATR. This shows that all the nodes were alive for a much longer time in LEAST compared to the other protocols.

\begin{figure}
    \centering
    \includegraphics[width = 0.8\columnwidth]{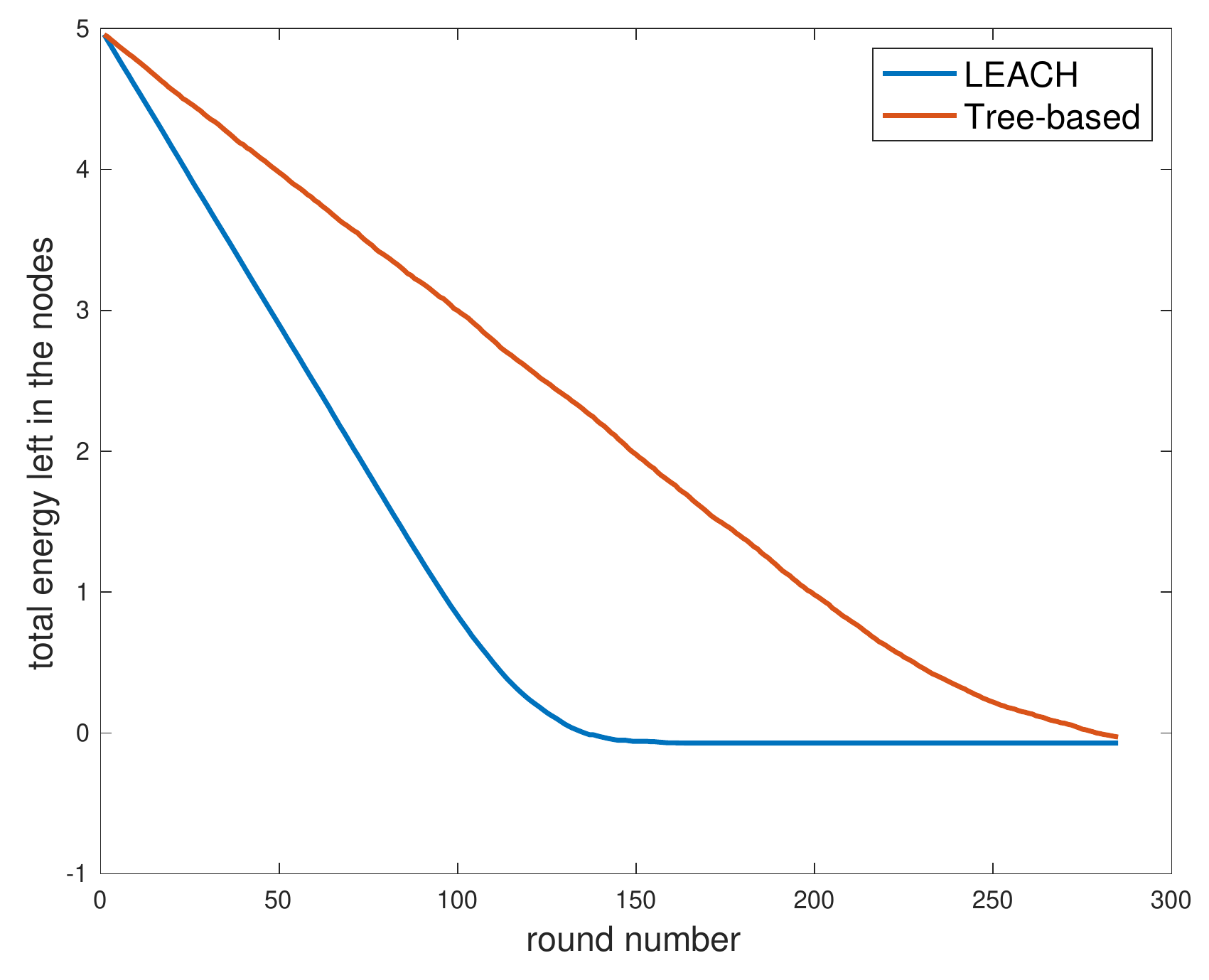}
    \caption{Comparison of LEAST with LEACH: total remaining energy per round}
    \label{energy}
\end{figure}

In Fig. \ref{energy}, the total energy remaining in all the sensor nodes during each round is plotted for both LEACH and LEAST. Two remarkable points can be spotted in this figure. Firstly, the energy of the network finishes much later for the LEAST protocol compared to the LEACH protocol. Additionally, the energy drop is much more steady for LEAST, whereas LEACH is exhibiting a sharper drop in the total energy level of the network. Overall, it can be observed in Fig. \ref{dead} and \ref{energy} that LEAST is outperforming LEACH and other protocols in terms of overall energy consumption of the network. 

\begin{figure}
    \centering
    \includegraphics[width = 0.8\columnwidth]{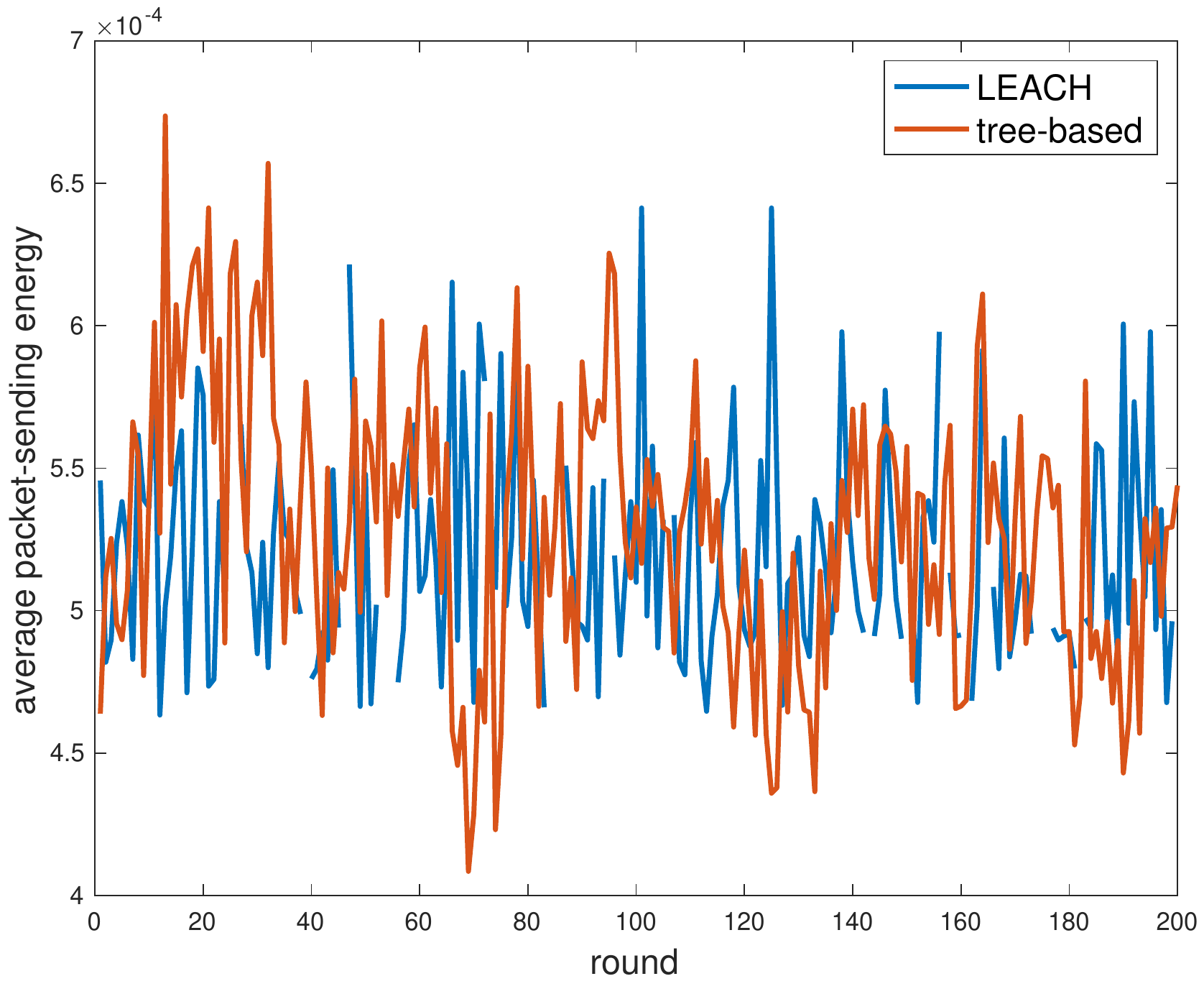}
    \caption{Comparison of LEAST with LEACH: average energy needed for sending a packet per node}
    \label{normal}
\end{figure}

It must be noted that the previous simulations only studied the setup phase of each round, which is dedicated to routing. In the steady-state phase, normal message packets will need to be transmitted in the network. To compare the energy spent for transmitting normal packets using the routing map of the network, we recreate the network using 30 nodes, randomly choose half of them in each round, and assume that each of them has one packet to send to the BS. Then, we average the energy spent by each node to make the transmission. Fig. \ref{normal} shows the result of this simulation. It can be seen that during the steady-state phase of both protocols, each node seems to be spending the same amount of energy on average. This essentially means that overall, LEAST saves energy, as it was shown that it is very energy efficient during the setup phase of the protocol.

\begin{figure}
    \centering
    \includegraphics[width = 0.8\columnwidth]{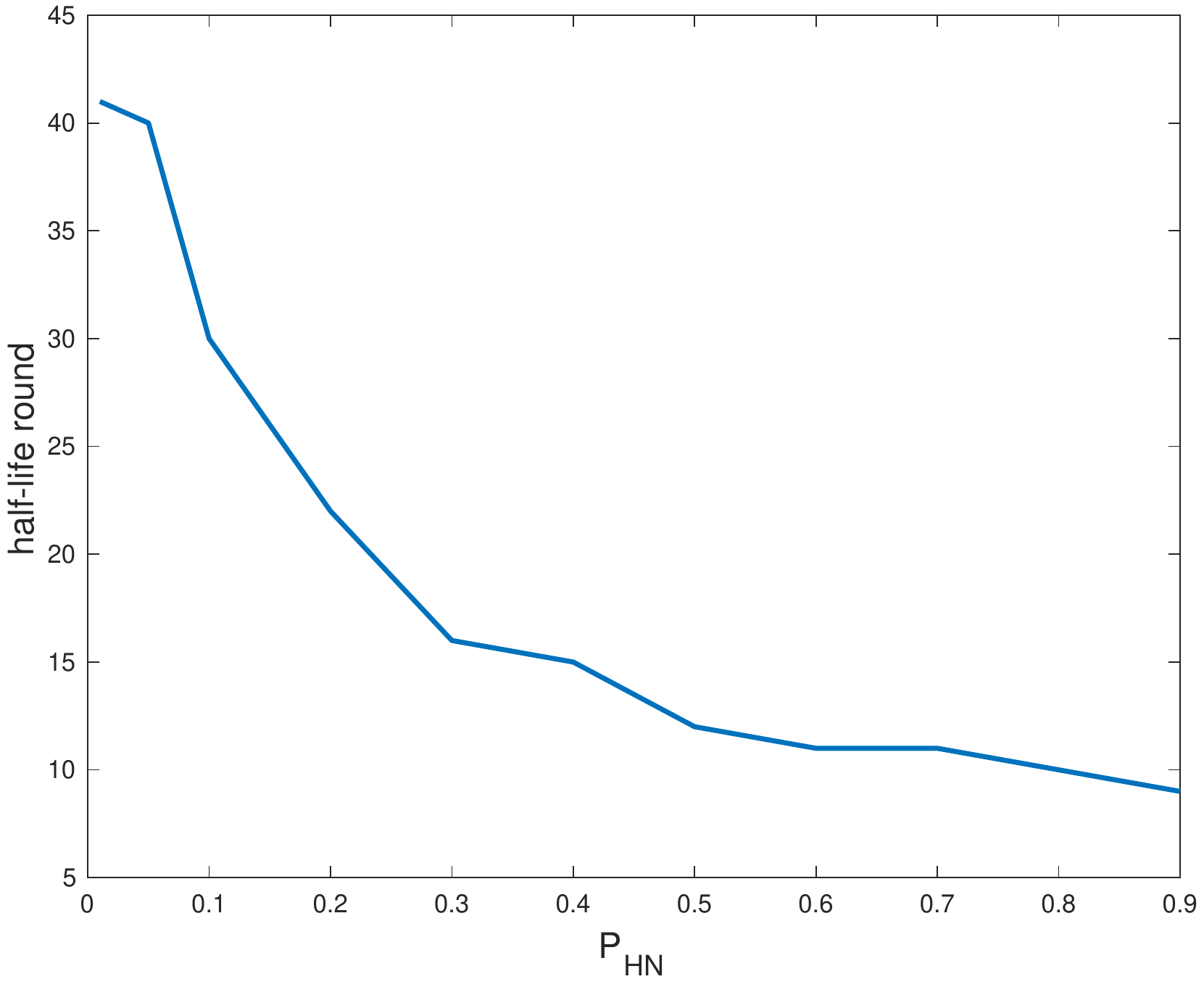}
    \caption{Effect of $p_{HN}$ on the lifetime of the network}
    \label{ph}
\end{figure}

Subsequently, we want to study the effect of the parameter $p_{HN}$ on the performance of the algorithm. For this purpose, we used the same setup as the first simulation, and tested the lifetime of the network for values of $p_{HN}$ between $0.01$ and $0.9$. Additionally, we reduced the initial energy of nodes to $0.005$ J so that the simulations run faster. Network lifetime is measured in terms of the number of rounds it takes for half of the nodes in the network to die, which we call half-life round. Fig. \ref{ph} shows the result of this simulation. It can be observed that this simulation suggests that lower values for $p_{HN}$ will result in a better performance. However, notice that choosing low values for $p_{HN}$ entails practical complications. This is because of the fact that the election process needs to repeat if no node is chosen as a host node, which causes delays and an increase in communication overhead.

\section{Conclusion and future work}

In this paper, LEAST was presented as a Low-Energy, Adaptive, Scalable and Tree-based routing protocol for Wireless Sensor Networks. It is adaptive as the routing map is a dynamic entity, which easily adapts to the distance between the sensors. Additionally, there are controlling parameters such $p_{CH}$, $p_{HN}$, and $p_H$, which can be specified according to the specific requirements and conditions of the network for which LEAST is going to be deployed. It is scalable as it is a hierarchical protocol, and the number of levels in the hierarchy of the network map does not have a limit. This means that as the number of the nodes increase and the distances grow, the network map scales to save transmission energy. For ease of deployment, LEAST was presented as an extension to LEACH. However, there are significant differences between LEAST and LEACH, namely the fact that the network map in LEAST can be a general tree, while that of LEACH is a three-level tree. Ultimately, It was shown that the energy consumption of LEAST was much lower than LEACH and other existing and well-known protocols. This means that using LEAST as the routing protocol for WSNs can result in a much longer lifetime for these networks.
\balance

Despite the fact that LEAST is currently in its elementary stages, it is still showing promising results and significant improvements over other WSN routing protocols. There are potential research directions that can be taken to improve LEAST. For instance, taking into account the remaining energy of the nodes in the process of relocating first-level nodes or choosing heir nodes can help improve the power consumption efficiency of LEAST. Additionally, tree compression algorithms such as \cite{farzaneh2022treeexplorer} can be used to reduce the communication overhead of transmitting the network map, which was shown to be a rooted tree. Furthermore, more research can be done on optimizing the network map to minimize distance between consecutive hops, the performance analysis of LEAST, and the practical aspects of deploying LEAST in WSNs.

\section*{Acknowledgment}

This work was supported by EPSRC grant number EP/T02612X/1. For the purpose of open access, the authors has applied a creative commons attribution (CC BY) licence (where permitted by UKRI, ‘open government licence’ or ‘creative commons attribution no-derivatives (CC BY-ND) licence’ may be stated instead) to any author accepted manuscript version arising. We also thank Moogsoft inc. for their support in this research project.

\bibliographystyle{ieeetr}

\begin{thebibliography}{10}

\bibitem{obaidat2014principles}
M.~S. Obaidat and S.~Misra, {\em Principles of Wireless Sensor Networks}.
\newblock Cambridge University Press, 2014.

\bibitem{heinzelman1999adaptive}
W.~R. Heinzelman, J.~Kulik, and H.~Balakrishnan, ``Adaptive protocols for
  information dissemination in wireless sensor networks,'' in {\em Proceedings
  of the 5th Annual ACM/IEEE International Conference on Mobile Computing and
  Networking}, pp.~174--185, 1999.

\bibitem{kulik2002negotiation}
J.~Kulik, W.~Heinzelman, and H.~Balakrishnan, ``Negotiation-based protocols for
  disseminating information in wireless sensor networks,'' {\em Wireless
  Networks}, vol.~8, no.~2, pp.~169--185, 2002.

\bibitem{intanagonwiwat2000directed}
C.~Intanagonwiwat, R.~Govindan, and D.~Estrin, ``Directed diffusion: A scalable
  and robust communication paradigm for sensor networks,'' in {\em Proceedings
  of the 6th Annual International Conference on Mobile Computing and
  Networking}, pp.~56--67, 2000.

\bibitem{sadagopan2003acquire}
N.~Sadagopan, B.~Krishnamachari, and A.~Helmy, ``The acquire mechanism for
  efficient querying in sensor networks,'' in {\em Proceedings of the First
  IEEE International Workshop on Sensor Network Protocols and Applications,
  2003.}, pp.~149--155, IEEE, 2003.

\bibitem{al2004routing}
J.~N. Al-Karaki and A.~E. Kamal, ``Routing techniques in wireless sensor
  networks: a survey,'' {\em IEEE Wireless Communications}, vol.~11, no.~6,
  pp.~6--28, 2004.

\bibitem{bhushan2019routing}
B.~Bhushan and G.~Sahoo, ``Routing protocols in wireless sensor networks,'' in
  {\em Computational Intelligence in Sensor Networks}, pp.~215--248, Springer,
  2019.

\bibitem{heinzelman2002application}
W.~B. Heinzelman, A.~P. Chandrakasan, and H.~Balakrishnan, ``An
  application-specific protocol architecture for wireless microsensor
  networks,'' {\em IEEE Transactions on Wireless Communications}, vol.~1,
  no.~4, pp.~660--670, 2002.

\bibitem{lindsey2002pegasis}
S.~Lindsey and C.~S. Raghavendra, ``Pegasis: Power-efficient gathering in
  sensor information systems,'' in {\em Proceedings, IEEE Aerospace
  Conference}, vol.~3, pp.~3--3, IEEE, 2002.

\bibitem{manjeshwar2001teen}
A.~Manjeshwar and D.~P. Agrawal, ``Teen: a routing protocol for enhanced
  efficiency in wireless sensor networks.,'' in {\em Proceedings of the 1st
  International Workshop on Parallel and Distributed Computing Issues in
  Wireless Networks and Mobile Computing}, vol.~1, p.~189, 2001.

\bibitem{li2001hierarchical}
Q.~Li, J.~Aslam, and D.~Rus, ``Hierarchical power-aware routing in sensor
  networks,'' in {\em Proceedings of the DIMACS Workshop on Pervasive
  Networking}, vol.~10, Citeseer, 2001.

\bibitem{kodali2014multi}
R.~K. Kodali and N.~K. Aravapalli, ``Multi-level leach protocol model using
  ns-3,'' in {\em 2014 IEEE International Advance Computing Conference (IACC)},
  pp.~375--380, Ieee, 2014.

\bibitem{ding2003aggregation}
M.~Ding, X.~Cheng, and G.~Xue, ``Aggregation tree construction in sensor
  networks,'' in {\em 2003 IEEE 58th Vehicular Technology Conference. VTC
  2003-Fall (IEEE Cat. No. 03CH37484)}, vol.~4, pp.~2168--2172, IEEE, 2003.

\bibitem{ma2014hybrid}
L.~Ma, H.~Leung, and D.~Li, ``Hybrid tdma/cdma mac protocol for wireless sensor
  networks,'' {\em Journal of Networks}, vol.~9, no.~10, p.~2665, 2014.

\bibitem{leachcodematlab}
M.~H. Homaei, ``Low energy adaptive clustering hierarchy protocol (leach).''
  \url{https://www.mathworks.com/matlabcentral/fileexchange/44073-low-energy-adaptive-clustering-hierarchy-protocol-leach},
  2022.
\newblock [Online; accessed September 09, 2022].

\bibitem{kim2005power}
H.-s. Kim and K.-j. Han, ``A power efficient routing protocol based on balanced
  tree in wireless sensor networks,'' in {\em First International Conference on
  Distributed Frameworks for Multimedia Applications}, pp.~138--143, IEEE,
  2005.

\bibitem{farzaneh2022treeexplorer}
A.~Farzaneh, M.-A. Badiu, and J.~P. Coon, ``Treeexplorer: a coding algorithm
  for rooted trees with application to wireless and ad hoc routing,'' {\em
  arXiv preprint arXiv:2207.05626}, 2022.

\end{thebibliography}

\end{document}